\documentclass[twocolumn,showpacs,preprintnumbers,amsmath,amssymb,latexsym,prd,footinbib,floatfix]{revtex4-1}

\usepackage{natbib}
\usepackage{amsmath}
\usepackage{graphicx}
\usepackage{relsize}
\usepackage{MnSymbol}
\usepackage{xcolor}
\usepackage{cleveref}
\usepackage[utf8]{inputenc}
\usepackage[pdf]{pstricks}

\begin{document}

\title{Radius of convergence in lattice QCD at finite $\mu_B$ with rooted staggered fermions}
\author{M. Giordano}
\author{K. Kapas}
\author{S.D. Katz}
\author{D. Nogradi}
\author{A. Pasztor}
\email[Corresponding author: ]{\url{apasztor@bodri.elte.hu}}

\affiliation{ELTE E\"otv\"os Lor\'and University, Institute for
  Theoretical Physics, P\'azm\'any P.\ s.\ 1/A, H-1117, Budapest,
  Hungary}

\begin{abstract}
In typical statistical mechanical systems the grand canonical partition 
function at finite volume is proportional to a polynomial of the fugacity 
$e^{\mu/T}$. The zero of this Lee-Yang polynomial closest to the origin 
determines the radius of convergence of the Taylor expansion of the 
pressure around $\mu=0$.
The computationally cheapest formulation of lattice QCD, rooted staggered 
fermions, with the usual definition of the rooted determinant, does not admit such a 
Lee-Yang polynomial. We show that the radius of convergence is then bounded by the spectral gap 
of the reduced matrix of the unrooted staggered operator.
This is a cutoff effect that potentially affects all estimates of the radius of convergence with the standard staggered rooting.
We suggest a new definition of the rooted staggered determinant at finite chemical potential that allows
for a definition of a Lee-Yang polynomial, and, therefore of the numerical study of Lee-Yang zeros. 
We also describe an algorithm to determine the Lee-Yang zeros and apply it to configurations 
generated with the 2-stout improved staggered action at $N_t = 4$. We perform 
a finite-volume scaling study of the leading Lee-Yang zeros and estimate 
the radius of convergence of the Taylor expansion extrapolated to an 
infinite volume. We show that the limiting singularity is not on the real 
line, thus giving a lower bound on the location of any possible phase 
transitions at this lattice spacing. In the vicinity of the crossover 
temperature at zero chemical potential, the radius of convergence turns 
out to be $\mu_B/T \approx 2$ and roughly temperature independent. 
Our simulations are performed at strange quark chemical potential $\mu_s=0$, but the 
method can be straightforwardly extended to strangeness chemical potential $\mu_S=0$ 
or strangeness neutrality.
\end{abstract}

\maketitle

\section{Introduction}

One of the open problems in the study of QCD at finite temperature and
density is determining the phase diagram of the theory in the
temperature ($T$)-baryon chemical potential ($\mu_B=3\mu_q$) plane. 
It is by now established that at $\mu_B=0$ there is an analytic 
crossover~\cite{Aoki:2006we,Bhattacharya:2014ara} at
a temperature~\cite{Aoki:2006br,Aoki:2009sc,Borsanyi:2010bp,Bazavov:2011nk} of $T_c\approx 150\text{--}160~{\rm MeV}$. It is further conjectured that in the 
$(T,\mu_B)$ plane there is a line of crossovers, departing from
$(T_c,0)$, that eventually turns into a line of first-order phase
transitions. The point $(T_{\rm CEP},\mu_{\rm CEP})$ separating
crossovers and first-order transitions is known as the critical
end point (CEP), and the transition is expected to be of second
order there. 

The conjectural phase diagram discussed above is mostly based on 
effective models of QCD~\cite{Fukushima:2013rx}. 
To settle the issue, one needs a first-principles study of QCD at
finite temperature and density,  
which requires nonperturbative tools like the lattice formulation of the
theory. Unfortunately, the introduction of a finite 
chemical potential $\mu_B$ makes the direct application of
traditional importance-sampling methods impossible due to the notorious sign
problem. For this reason, lattice QCD could so far give
very limited information about the phase diagram away from $\mu_B=0$.

While a solution of the sign problem is still lacking, several
techniques have been developed to bypass it, including Taylor
expansion at $\mu_B=0$~\cite{Gavai:2003mf,Gavai:2004sd,Allton:2005gk,Gavai:2008zr,
Basak:2009uv,Borsanyi:2011sw,Borsanyi:2012cr,Bellwied:2015lba,
Ding:2015fca,Bazavov:2017dus}, analytic continuation from imaginary
chemical potential~\cite{deForcrand:2002hgr,DElia:2002tig,DElia:2009pdy, 
Cea:2014xva,Bonati:2014kpa,Cea:2015cya,Bonati:2015bha,Bellwied:2015rza,DElia:2016jqh,Gunther:2016vcp,Alba:2017mqu,
Vovchenko:2017xad,Bonati:2018nut,Borsanyi:2018grb}, and reweighting
methods~\cite{Hasenfratz:1991ax,Barbour:1997ej,Fodor:2001au,Fodor:2001pe,Allton:2002zi,Fodor:2004nz,Csikor:2004ik,Ejiri:2005ts}.  
The basic idea of these methods is to reconstruct the behavior of the
theory at finite real chemical 
potential, where standard simulations are not feasible, by
extrapolating from zero or purely imaginary chemical potential, where
the sign problem is absent. A problem of all methods of this
type is the overlap problem, i.e., the incorrect sampling of the
important configurations of the system, which becomes exponentially
severe as the volume of the system increases.

Among these methods, reweighting has the advantage of having no other
systematic error besides 
the overlap problem and in
principle would lead to the correct results in the limit of infinite
statistics, so that it can be at least used as a ``brute-force''
approach to the sign problem. On the other hand, the Taylor
expansion method is also affected by 
systematic errors from the truncation of the Taylor series, as well
as the existence of a finite radius of convergence, while
extrapolation from imaginary chemical potential involves a 
rather 
uncontrolled analytic continuation in $\mu_B$. 

The aim of this paper is to obtain as much information as possible about
the analytic structure of the pressure in the complex baryon chemical potential plane.
In particular, we will estimate the position of the singularity closest to $\mu_B=0$, 
which provides both a lower bound on the location of possible phase transitions 
on the phase diagram as well as the limit of reliability of the equation of state
coming from a Taylor expansion, which is an important input for the phenomenology of heavy ion collisions.
 
In a finite volume the analytic properties of the pressure of a typical statistical mechanical system
are governed by the zeros of the partition function in the complex-$\mu_B$
plane, the so-called Lee-Yang zeros~\cite{Lee:1952ig}. In general,
the grand-canonical partition function of a relativistic lattice system at
finite $\hat{\mu}=\mu_q/T=\mu_B/{3T}$ in a finite spatial volume $V$
is a polynomial in the fugacity $z=e^{\hat{\mu}}$, which we may call
the Lee-Yang polynomial, times a nonvanishing factor
$e^{-kV\hat{\mu}}$ for some model-dependent constant $k$. The Lee-Yang 
zeros are the singular points of the pressure, $p=-\frac{T}{V}\log Z$, as a
function of complex fugacity or chemical potential. The accumulation
of such zeros near the real $\mu_B$ axis in the thermodynamic limit
$V\to\infty$ signals the presence of a genuine phase transition. In
the case of a crossover, no nonanalyticity develops on the real line
and the distance of the Lee-Yang zeros  from the real axis provides a
measure of the strength of the transition. 
The radius of convergence of the Taylor expansion of the pressure 
around $\mu_B=0$ is equal to the distance from the origin 
of the Lee-Yang zero closest to it, which we will refer to as the leading 
Lee-Yang zero. Depending on the behavior of the leading zero in the infinite-volume 
limit, the radius of convergence could correspond to the chemical potential at
which an actual phase transition takes place (in case that the
imaginary part of the leading zero extrapolates to zero) or just give
a lower bound on the location of a phase transition (in case that the
imaginary part extrapolates to a nonzero value).  

Conversely, the position of the closest Lee-Yang zero can be
inferred from the high-order behavior of the Taylor
coefficients. In fact, at
a fixed lattice spacing, as long as one uses a discretization where the 
partition function is an entire function of $\mu$, knowing the position 
of the leading Lee-Yang zero is completely equivalent to knowing 
the asymptotically high-order behavior of the Taylor coefficients of the pressure. 
This was shown in~\cite{Giordano:2019slo}, where
explicit formulas are given for the conversion~\footnote{We note here that when converting the 
high order Taylor coefficients to the distance 
of the closest Lee-Yang zero, i.e. the radius of convergence,
one must not use the ratio estimator, since that estimator is
guaranteed to not converge in any finite volume. This was also 
proved in~\cite{Giordano:2019slo}. Unfortunately, the use of the 
ratio estimator is nevertheless far spread in the literature}.
The determination of the leading Lee-Yang zero from the Taylor
coefficients however involves an extrapolation to high orders of the
Taylor expansion, which is technically challenging. 
It turns out that a direct determination of the Lee-Yang zeros using
reweighting techniques is instead more straightforward.
One might wonder how this is possible: if it is so difficult to
estimate reliably the high-order Taylor coefficients in 
order to determine the radius of convergence, determining the latter
directly seems hopeless. The answer is that the large (above 100\%)
errors on the high-order coefficients are strongly correlated, and
cancel out in the particular combinations that give the leading 
Lee-Yang zero, and therefore the radius of convergence. 
This surprising conclusion was discussed in
Ref.~\cite{Giordano:2019slo} and demonstrated explicitly in a
numerical study of unrooted staggered fermions on a small lattice.
This suggests that it is more efficient to calculate the radius
of convergence first at a finite lattice spacing, where the 
strong correlations between the Taylor coefficients are present,
than taking the continuum limit of the coefficients first and
calculating the radius of convergence of the continuum expansion,
in which case part of the correlations is lost.

A convenient way to do reweighting, which allows for a straightforward
determination of the Lee-Yang zeros, is to compute the spectrum of the
so-called reduced matrix
$P$~\cite{Hasenfratz:1991ax,Fodor:2001pe,Fodor:2004nz} on an ensemble  
of gauge configurations at $\mu_q=0$, which then allows one to reweight to
any finite $\mu_q$ using the relation $\det M(\hat{\mu}) = e^{-kV\hat{\mu}}\det(P -
e^{\hat{\mu}})$, where $M(\hat{\mu})$ is a lattice discretization of the
QCD Dirac operator at finite $\mu_q$ and $k$ is the same 
model-dependent constant appearing in the partition function.
The reduced matrix of Ref.~\cite{Hasenfratz:1991ax} is instrumental to the approach of this
paper. In fact, expressed in terms of the reduced matrix, the fermionic determinant is
a polynomial in fugacity on each configuration 
up to a nonvanishing prefactor, which allows for a straightforward reconstruction
of the polynomial part of the grand-canonical partition function at
finite $\mu_q$ and the subsequent determination of the Lee-Yang zeros
by means of standard numerical techniques.

The discussion above is quite general, and it applies whenever a
reduced matrix formulation is available~\cite{Hasenfratz:1991ax,Danzer:2008xs,Alexandru:2010yb}. 
Unfortunately, this does
not include the computationally most convenient formulation of lattice 
fermions, namely the rooted staggered discretization. 
Near the continuum the spectrum of the staggered Dirac operator 
shows quartets of near-degenerate eigenvalues, with relative
splittings of order $O(a)$, corresponding to the so-called staggered
tastes~\cite{Follana:2004sz,Follana:2005km,Durr:2004as}.  
Taking roots of the staggered determinant should then fully solve
the doubling problem of lattice fermions, by reducing the $N_f=4$
flavor theory of unrooted staggered fermions down to the desired
number of degenerate flavors. In the $N_f=2$ case, for example, the
square root of the determinant is taken. This procedure has become
standard, and although there is no rigorous proof that it ultimately
provides us with a genuine local continuum quantum field theory, 
the results obtained are in good agreement with experiments 
and with lattice 
results obtained using other fermion discretizations~\cite{Borsanyi:2012uq,Bhattacharya:2014ara,Borsanyi:2015waa,Borsanyi:2016ksw,Borsanyi:2015zva}. 
At finite chemical potential one can still recast the determinant of
the staggered Dirac operator in terms of a reduced matrix, but the
rooted determinant is not a polynomial in fugacity anymore, and so
the essentially polynomial character of the grand-canonical partition
function is lost. 

The lack of a Lee-Yang polynomial is not the only problem afflicting
rooted staggered fermions at finite $\mu_B$. Unlike at $\mu_B=0$, where
one can simply take the real positive root of the real positive
determinant, at finite $\mu_B$ where the determinant is complex there is
not such a natural choice, and some other criterion is needed to
resolve the intrinsic ambiguity of rooting. The standard choice for
both reweighting and the Taylor-expansion method is to take the root
that, on a given gauge field configuration and as a function of $\mu_B$,
continuously connects to the real positive root at $\mu_B=0$. 
While this choice is perfectly fine in the continuum limit, where
the formation of eigenvalue quartets is expected to make the
ambiguities related to rooting go away, at any finite spacing it
leads to serious analyticity problems. As we will argue
the resulting partition function as a
function of complex fugacity will be nonanalytic everywhere on the
support of the spectrum of the reduced matrix $P$. 
The radius of convergence of the Taylor expansion of the pressure will
then be given by the spectral gap of $P$, i.e. the closest distance where the
spectral density of $P$ is nonzero, which on a finite ensemble corresponds to the distance of
the closest eigenvalue on the entire ensemble. This spectral gap a priori 
has nothing to do with Lee-Yang zeros or phase
transitions. This means the radius of convergence 
will not be given by a partition function zero, unless it happens to be inside the gap. Even if this lucky coincidence happens, one would not be able to
tell from the Taylor expansion. 

We will circumvent this problem by
defining the ``rooted'' staggered determinant not by choosing a
particular branch of the root function, but by setting it equal to a
polynomial in the fugacity that 
is expected to converge to the same continuum limit as the standard
procedures discussed above. Such a polynomial will differ from the
standard definitions of the rooted determinant by terms that are
nonanalytic in $\mu_B$ but vanish in the continuum. This way the radius
of convergence of the expansion of the pressure will certainly not be
related anymore to the spectral gap (which depends on the support of
the spectrum on the whole ensemble and is thus determined by some
``extreme'' configuration) and will be given by an actual Lee-Yang
zero (which depends only on averages over the ensemble).

Our construction of a rooted determinant in polynomial form is
motivated by the very idea behind rooting of staggered fermions, i.e.,
the formation of taste quartets in the continuum. In
Ref.~\cite{Golterman:2006rw} it was suggested that the optimal way to
do rooting is to identify the taste multiplets in the spectrum of the 
staggered Dirac operator
and replace them with their average to define the rooted
determinant. This was argued to considerably reduce the finite-spacing 
effects compared to other procedures. As we will argue, the formation
of quartets in the spectrum of the staggered Dirac operator leads to
the formation of quartets in the spectrum of the corresponding reduced
matrix $P$. In the spirit of Ref.~\cite{Golterman:2006rw}, we can
therefore define a rooted determinant in the reduced matrix approach
by judiciously grouping the eigenvalues of $P$ and replacing them with
a properly defined average. In this way we automatically obtain a
definition of the rooted determinant which is a polynomial in
fugacity and so analytic in $\mu_B$. One can then define a Lee-Yang
polynomial and carry out the study of its zeros by standard methods.
The purpose of this paper is thus twofold: i) show how to conveniently
group eigenvalues and how to average them in order to provide a more convenient definition 
of the theory at finite $\mu_B$ with two light flavours of rooted staggered fermions; ii)
obtain a direct and complete determination of the Lee-Yang zeros of
the so-defined partition function. 

The plan of the paper is the following. In Sec. II we review the
reduced-matrix approach, focusing in particular on its problems of
analyticity at finite $\mu_B$ when one takes roots of the determinant,
and how to solve these problems by grouping eigenvalues. We also
briefly review Lee-Yang zeros and discuss how to numerically compute
them. In Sec. III we perform a numerical study with 2-stout 
improved $N_t=4$ staggered lattices. Finally, in Sec. IV we draw our 
conclusions and discuss future prospects. 

\section{Reduced matrix for staggered fermions}
\label{sec:redmat}

\subsection{Generalities}
\label{sec:general}

The introduction of a finite quark chemical potential $\mu=\mu_q$ in the 
staggered Dirac operator is usually done by coupling it to the
temporal links as follows:
\begin{equation}
  \label{eq:stag}
  \begin{aligned}
 D_{\rm stag}(a\mu) &=  \textstyle\frac{1}{2}\eta_4\left[ e^{a\mu} U_4 T_4
      -  e^{-a\mu} T_4^\dag U_4^\dag \right] + D_{\rm stag}^{(3)}\,,\\
    D_{\rm stag}^{(3)} &= \textstyle\frac{1}{2}\sum_{j=1}^3\eta_j\left[ U_j T_j
      -   T_j^\dag U_j^\dag \right]\,,
  \end{aligned}
\end{equation}
where $a$ is the lattice spacing, $(T_\alpha)_{xy} =
\delta_{x+\hat{\alpha},y}$ are the translation operators and
$\eta_\alpha$ are the usual staggered phases. It is easy to show that
$\det M(\hat{\mu}) \equiv \det \left[ 2(D_{\rm stag}(a \mu) + am) \right]$ depends only on
$\hat{\mu}=\mu/T$. It has been shown in Ref.~\cite{Hasenfratz:1991ax}
that~\footnote{The factor of 3 in the exponential before the product
  is missing in Ref.~\cite{Hasenfratz:1991ax}.} 
\begin{equation}
\label{eq:stagred}
\det M(\hat{\mu}) = e^{3 V \hat{\mu} } \prod_{i=1}^{6 V} \left(
  \xi_i - e^{\hat{-\mu}} \right)
\,,
\end{equation}
where $V= N_s^3$ is the spatial volume and $N_s$ is the spatial linear
size of the lattice in lattice units (which must be an even number), 
and $\xi_i$ are eigenvalues of the reduced matrix $P$.  
In the temporal gauge [$U_4(t,\vec{x})=1$ for $0\le t < N_t-1$], this reads 
\begin{equation}
\label{eq:stagred2}
  \begin{aligned}
	P&=-\left(\prod_{i=0}^{N_t-1}P_i\right) L \,,  &&&
P_i&=   \begin{pmatrix} B_i & 1 \\ 1 &
    0 \end{pmatrix}	\,, \\
B_i &= 2\eta_4( D^{(3)} + am)\big|_{t=i}\,,
&&& L &=
\begin{pmatrix}
  U_4 & 0 \\ 0 & U_4
\end{pmatrix}\Big|_{t=N_t-1}\,,
  \end{aligned}
\end{equation}
i.e., $\eta_4 B_i$ is the sum of the spatial derivatives and mass
parts of the staggered matrix on the $i$th time-slice, and $L$ is the
block-diagonal matrix of temporal links on the last time slice
(i.e., the untraced Polyakov loops). Since $P$ is $\mu$ independent,
knowledge of the $\xi_i$ for a given gauge configuration allows to
compute the corresponding unrooted quark determinant for arbitrary
$\mu$. From a Monte Carlo simulation at $\mu=0$ one can then obtain
the grand-canonical partition function at any $\mu$ via
\begin{equation}
  \label{eq:mupartfunc}
  \begin{aligned}
      {\cal P}_\xi(e^{\hat{\mu}}) &\equiv \prod_{i=1}^{6V} \frac{\xi_i - e^{\hat{\mu}}}{ \xi_i - 1 } \\
      {\cal P}(e^{\hat{\mu}})     &\equiv \left\langle {\cal P}_\xi(e^{\hat{\mu}}) \right\rangle_0 \\
      \frac{Z(\hat{\mu})}{Z(0)}          &= \left\langle  \frac{\det M(\hat{\mu})}{\det M(0)} \right\rangle_0 = e^{-3 V\hat{\mu} }\,{\cal P}(e^{\hat{\mu}}) \,,
  \end{aligned}
\end{equation} 
where the subscript 0 indicates that the expectation value is computed
at $\mu=0$. The quantities ${\cal P}_\xi$ and ${\cal P}$, defined in Eq.~\eqref{eq:mupartfunc}, are
polynomials of degree $6V$ of the fugacity $z=e^{\hat{\mu}}$. The
coefficients of the Lee-Yang polynomial ${\cal P}$ are the average of
those of ${\cal P}_\xi$, and coincide with the (normalized) canonical
partition functions, 
and as such they are positive quantities. Notice that Roberge-Weiss
symmetry~\cite{Roberge:1986mm} imposes that only coefficients of order
$3n$ can be nonzero. The canonical partition functions are usually
obtained as the coefficients of a Fourier expansion of the 
grand-canonical partition function at imaginary chemical
potential~\cite{Hasenfratz:1991ax,Morita:2012kt,Fukuda:2015mva,Nakamura:2015jra,deForcrand:2006ec,Ejiri:2008xt,Li:2010qf,Li:2011ee,Danzer:2012vw,Gattringer:2014hra,Boyda:2017lps,Goy:2016egl,Bornyakov:2016wld,Boyda:2017dyo,Wakayama:2018wkc,Nagata:2014fra}. 
Our direct determination from the eigenvalues of $P$ is free from the
systematic uncertainty associated with the extraction of Fourier
coefficients from a discrete set of imaginary chemical potentials.

The matrix $P$ has a few nice properties, which we list here:
\begin{enumerate}
\item $\det P = 1$.
\item Its eigenvalues come in pairs $(\xi_i,1/\xi_i^*)$.
\item The product $\sideset{}{'}\prod_i \xi_i$ of eigenvalues inside
  the unit circle is real positive.
\end{enumerate}
A proof of these properties can be found in the Appendix. 

The Lee-Yang zeros are the roots of the polynomial ${\cal P}(e^{\hat{\mu}})$.
Due to the symmetries of the partition function, the Lee-Yang zeros
will be symmetric under the reflection $\mu \to -\mu$ due to $CP$
symmetry and under the reflection $\mu \to \mu^*$ due to the fact that
$Z(\hat{\mu})$ is real analytic. Furthermore, Roberge-Weiss symmetry
implies that we can restrict ourselves to the strip  ${\rm Im}\,\hat{\mu}\in 
[-\frac{\pi}{3},\frac{\pi}{3}]$, since the zeros are then repeated 
with a period of $\frac{2\pi}{3}$ in the imaginary $\hat{\mu}$ direction. 
There can be no zeros on the real axis due to positivity of $Z(\mu)$ and on
the imaginary axis since the determinant is real positive there.
Note that these properties of $Z$ originate in properties 1--3) of $P$.

\subsection{Problems with rooting}
\label{sec:probroot}

When using rooted staggered fermions, one must replace the ratio of
determinants in Eq.~\eqref{eq:mupartfunc} with its appropriate root. The
intrinsic ambiguity associated with rooting is easily solved at
$\mu=0$, where it is natural to take the real positive root of the
real positive ratio of determinants. Such a natural possibility is not
available at nonzero $\mu$ where the fermionic determinant is
generally complex. 
The choice in previous reweighting studies~\cite{Fodor:2001pe} was to use 
the root that on a fixed gauge configuration continuously connects to the
real positive root as a function of $\mu$. We focus here on the
$N_f=2$ case.
In terms of the eigenvalues of the reduced matrix one has
\begin{equation}
  \label{eq:oldrooting}
    \sqrt{\frac{\det M(\hat{\mu})}{\det M(0)}}_{\rm rew} \equiv e^{+\frac{3}{2}V \hat{\mu}} \prod_n\sqrt{\frac{\xi_n-e^{-\hat{\mu}}}{\xi_n-1}}\rm{,}
\end{equation}
with the branch cut of the square root chosen to lie along the negative real axis. 
Note that this equation actually provides a definition of the square
root on the left-hand side. 
On a single configuration, this quantity has $6V$ branch cuts parallel to the real axis. 
Restricting to real chemical potentials one always moves parallel to
the branch cuts and never crosses them, making the function continuous at real $\mu$.
The Taylor-expansion method in turn is formulated by taking the square root of the entire ratio of determinants $\frac{\det M(\hat{\mu})}{\det M(0)}$, 
and expanding around $\mu=0$, where the argument of the square-root function is one. 
This leads to an expansion
\begin{equation}
  \label{eq:Taylor_rooting}
	\begin{aligned}
    \log \sqrt{\frac{\det M(\hat{\mu})}{\det M(0)}}_{\rm Taylor}
    &\equiv \sum_{n=1}^{\infty}{\frac{1}{2} \frac{\partial^n \log \det
        M(\hat{\mu})}{\partial \hat{\mu}^n} }\Big|_{\mu=0} \hat{\mu}^n
	\end{aligned}
\end{equation}
On a single configuration the radius of convergence of this Taylor
expansion is given by the point where the unrooted determinant first
becomes zero, i.e., by the closest eigenvalue of 
the reduced matrix. Notice that one would get the same formula if one
expanded $\log \sqrt{\frac{\det M(\hat{\mu})}{\det M(0)}}_{\rm rew}$
around zero, meaning that on a single configuration the two methods
define the same rooted determinant within the radius of convergence. 

The discussion above extends naturally to the Taylor expansion of
$\log Z$ around $\mu=0$, but now the radius of convergence is
determined by the eigenvalue of the reduced matrix 
whose logarithm is closest to $0$ on the entire ensemble.
The issue is easily understood with a very simple example, in
which we have only two configurations and two eigenvalues on each
configuration. In this case the toy partition function is defined as 
\begin{equation}
  \label{eq:toyZ}
    Z_{\rm toy}(\zeta) = \sqrt{\zeta-a_1} \sqrt{\zeta-a_2} + \sqrt{\zeta-b_1} \sqrt{\zeta-b_2}\,,
\end{equation}
where $a_{1,2},b_{1,2}$ are complex numbers, the square roots are all
defined with the branch cut on the negative real axis
and $\zeta$ is the complex fugacity parameter $\zeta=e^{\hat{\mu}}-1 =\hat{\mu} + O(\hat{\mu}^2)$. 
No matter how close $a_1,a_2$ and $b_1,b_2$ are, 
the radius of convergence of the Taylor expansion of the ``pressure''
$\log Z_{\rm toy}$ around $\zeta=0$ cannot be larger than
$\min(|a_1|,|b_1|,|a_2|,|b_2|)$, leading to our statement that the
radius of convergence is bounded from above by the spectral gap. This
fact remains true for any $a_1 \neq a_2$, $b_1 \neq b_2$, and only
changes in the case of exact degeneracy, which in this toy example
mimics the ``continuum limit''. 
The partition function $Z_{\rm toy}$ can still have a zero, that in general can be either closer or farther away 
from $\zeta=0$ than the closest square-root branch point. 
In the limit of exact degeneracy, $a_1=a_2:=a$ and $b_1=b_2:=b$ the radius 
of convergence of $\log Z_{\rm toy}$ is given by the partition function zero at $(a+b)/2$. 

\begin{figure}[t]
  \centering
  \rotatebox{-90}{\includegraphics[width=0.33\textwidth]{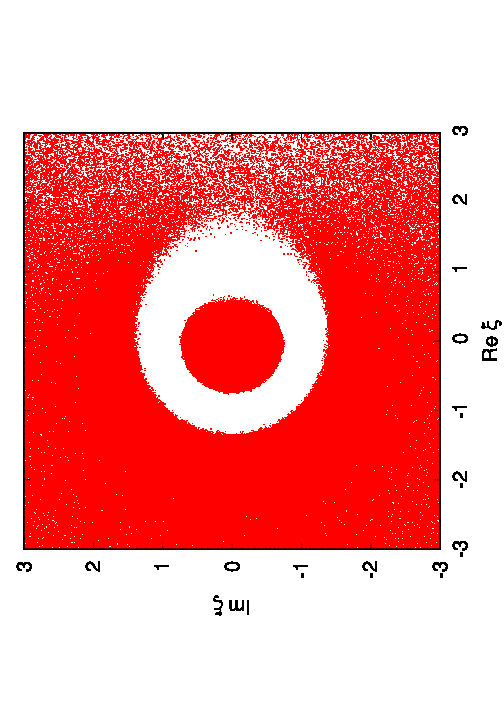}}
  \caption{Eigenvalues of the reduced matrix $P$ close to the unit circle on $3000$ configurations at a lattice volume of $12^3 \times 4$ and $\beta=3.35$ superimposed into a single plot. 
    A small gap is also present near $0$, but it is not visible on the plot.}
  \label{fig:single_conf}
\end{figure}

The situation is not expected to improve if one has a larger number 
of branch points in the square root and if instead of the sum of two terms one has
an average over the position of the branch points with some
probability distribution, which is the general case to which the
rooted fermion determinant in the reduced-matrix approach belongs.
In this case one expects that after averaging/integration over gauge
fields, the partition function will be nonanalytic over the support
of the spectrum of $P$. In fact, the only way to avoid this conclusion is the existence of
some cancellation mechanism between the branch points, which seems
unlikely. This is problematic not only for the
reduced-matrix approach, but for any approach involving the rooting of
the fermionic determinant, like the Taylor-expansion method or
analytic continuation from imaginary chemical potential. In fact, this
argument suggests that since the eigenvalues in the full gauge
ensemble are expected to fill densely some region inside the unit
circle (see Fig.~\ref{fig:single_conf}), at any finite spacing there will be an
analytically inaccessible region, whose  
boundary is determined by the extreme edges of the spectrum. 
Numerical results seem to indicate the existence of a gap
around the unit circle and thus of a finite domain of analyticity~\cite{Gibbs:1986hi,Fodor:2007ga,Nagata:2012mn}.

From a practical point of view, with the usual definitions of the
rooted determinant, the analyticity domain of the partition function
in fugacity on any finite ensemble of configurations will be
determined by the position of the eigenvalue of $P$ closest to $1$ in the ensemble.

A consequence of this discussion is that with the usual definition of the
rooted staggered determinant at finite $\mu$, the radius of convergence 
at a finite lattice spacing may not be related to the physics of phase 
transitions.
This also means that the radius of convergence of the continuum Taylor series
may not be equal to the continuum limit of the finite-spacing radius of convergence.

\begin{figure}[t]
  \centering
  \rotatebox{-90}{\includegraphics[width=0.33\textwidth]{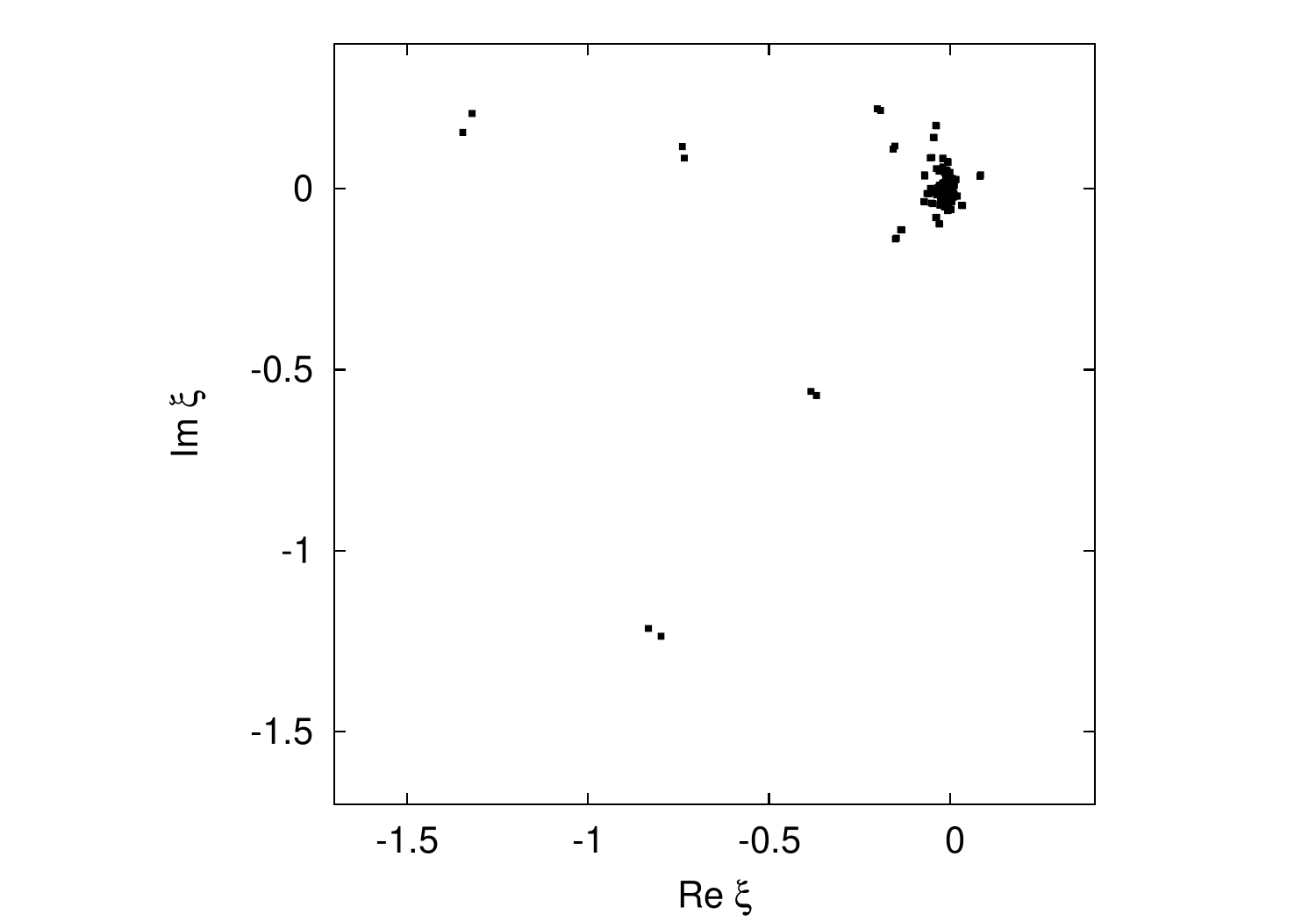}}
  \caption{Eigenvalues of $P$ after 50 stout smearing steps with $\rho=0.05$ 
    on a single gauge configuration on a $6^3\times 4$ lattice. The
    formation of doublets of eigenvalues is apparent.}
  \label{fig:smear}
\end{figure}

Before discussing the solution to these problems in the next
subsection, it is worth remarking that the possible easing of problems
near the continuum limit is based on the expected formation of
quartets of eigenvalues. So far, the formation of quartets near the
continuum limit has been investigated only for the usual staggered
operator~\cite{Follana:2004sz,Follana:2005km,Durr:2004as}. It is then
worth checking that they do indeed form also in the spectrum of the
reduced matrix. A first check is provided by solving analytically the
eigenvalue problem in the free case, in which quartets of eigenvalues
are explicitly shown to appear (see the Appendix). Since in the continuum
limit the relevant configurations fluctuate around the free one,
quartets of eigenvalues are expected to show up for sufficiently small
lattice spacing. We also conducted a second, numerical check.
While a direct study is currently out of reach, here
we mimic the approach to the free case by applying a large
number of stout smearing steps to a fixed gauge configuration on a
small $6^3\times 4$ lattice. After 50 steps of stout smearing~\cite{Morningstar:2003gk} with
$\rho=0.05$ doublets appear (see Fig.~\ref{fig:smear}), while a much
larger number of smearing 
steps ($\sim 150$) is needed for the appearance of quartets. This
behavior matches what has been observed with the usual staggered
operator~\cite{Durr:2004as}. Ultimately, one will have to go to fine 
lattices at a fixed temperature, and look at the spectrum of $P$ to
make sure there really is a quartet structure.

\subsection{Geometric matching}
\label{sec:geomatch}

Besides the analyticity problems discussed in the previous subsection,
rooting of the fermion determinant in the reduced-matrix approach is
expected to introduce systematic finite-spacing effects analogous to
those discussed in Ref.~\cite{Golterman:2006rw} for the usual
fermionic matrix. This is because of the existence of configurations where
the taste multiplets are cut through by the branch cut on the negative
real axis. While a sensible rooting procedure should correspond in
practice to replacing a multiplet with a single effective eigenvalue 
at the multiplet position, taking roots eigenvalue by eigenvalue as in Eq.~\eqref{eq:oldrooting}
can lead to the effective eigenvalue being far away, having the
opposite sign. 

The solution to both kind of problems is to identify
taste quartets of eigenvalues and replace them by a properly defined
average. In principle this could be done also on not so fine lattices
by, e.g., tracking the eigenmodes as the gauge configuration undergoes
a number of smearing steps, seeing which ones end up forming quartets,
and grouping them accordingly. This procedure is computationally very
expensive, and so one would rather opt for the next best thing: find
the group of closest eigenvalues and treat them as if they were taste
quartets. This will eventually become equivalent to the optimal
procedure in the continuum limit, since the near-continuum taste
quartets are well separated from each other and will automatically be
identified as the groups of closest eigenvalues. After identifying the
quartets, they are replaced by the fourth power of an appropriate
average, and the rooted determinant is obtained by retaining a single
power. 

Before detailing the procedure for actual lattice QCD let us come back
briefly to our toy example of the previous subsection and see how
this approach improves the analyticity of the partition function. 
Using the geometric mean as the
average within each pair, we define the rooted toy partition function as
\begin{equation}
    Z_{\rm{toy,poly}} = (\zeta-\sqrt{a_1 a_2}) + (\zeta-\sqrt{b_1 b_2})\,.
\end{equation}
The partition function is now a polynomial, and the radius of
convergence of $\log Z_{\rm{toy,poly}}$ around $0$ is determined by
its Lee-Yang zero $\frac{\sqrt{a_1 a_2}+\sqrt{b_1 b_2}}{2}$, which
continuously tends to $\frac{a+b}{2}$ as the splitting is diminished.  

Extending the toy example to the case of the lattice QCD partition
function, estimated using $N_{\rm{conf}}$ gauge field configurations
and computing the $6V$ eigenvalues of the reduced matrix, with the
usual rooting procedure we expect to find $6V N_{\rm{conf}}$ branch
points of square-root type, and no easy way to count the zeros of the
partition function. This also means the presence of $6V N_{\rm{conf}}$
square-root type branch points in $\log Z$, besides the logarithmic ones
originating from the zeros. Using a matching procedure to replace pairs of
eigenvalues with their geometric mean one finds instead no branch
points in $Z$, since the resulting partition function is a polynomial of order
$3V$, thus with exactly $3V$ zeros. This also means that $\log Z$ has
only logarithmic singularities, and that the radius of convergence of
its expansion around $\mu=0$ is determined by the closest one of them.

We now give details on how the matching procedure is actually
implemented in practice. The guidelines are the following:
\begin{itemize}
\item The correct quartets must be automatically selected in the
  continuum.
\item The new eigenvalues should retain the
  properties 1--3 of the unrooted reduced matrix
  discussed at the end of Sec.~\ref{sec:general}, in order to retain
  properties of the partition function $Z$ itself.
\end{itemize}
In this way we expect that in the continuum limit one finds a
well-defined continuum theory of a single fermion flavor with the
desired properties. Since we are interested in the case $N_f=2$ of two
light flavors, we can simplify the procedure and content ourselves
with identifying doublets of eigenvalues instead of quartets. To make
our proposal concrete, we still have to specify how to identify
doublets and how to average them.

\begin{figure}[t]
  \centering
    \rotatebox{-90}{\includegraphics[width=0.33\textwidth]{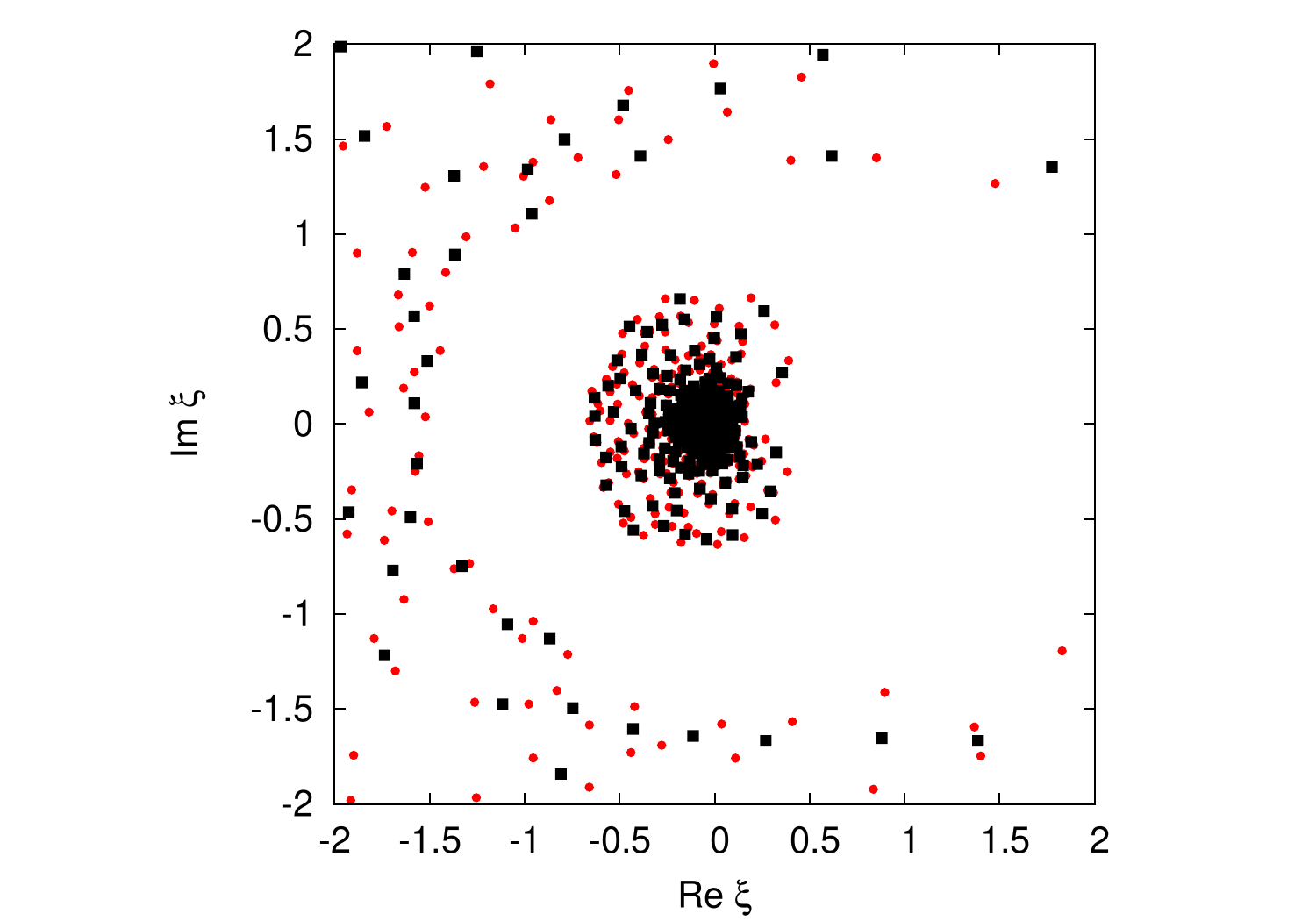}}
    \caption{The original eigenvalues of $P$ (red circles) together with the new eigenvalues after the
	         rooting via geometric matching (black boxes) for a $12^3 \times 4$ lattice at $\beta=3.35$.}
  \label{fig:pairs}
\end{figure}

The identification of doublets is achieved by minimizing the total sum
of the distances within each pair, which we dub ``geometric
matching''. As already remarked above, this will provide the correct
identification in the continuum limit, where nearly degenerate taste 
quartets are well separated from each other (except possibly for UV
related modes that do not affect the long-distance physics).
The relevant optimization problem can be solved in polynomial time by
means of the so-called Blossom algorithm~\cite{Edmonds}. A publicly
available implementation of this algorithm exists~\cite{Blossom},
which we used in our analysis. 

Once doublets $(\xi_1,\xi_2)$ of eigenvalues of $P$ are identified,
their product is replaced by their geometric mean, leading to the
desired rooted determinant. This procedure is illustrated in
Fig.~\ref{fig:pairs}. 
More precisely, we replace $\xi_1 \xi_2 \to \tilde{\xi}$ obtained as follows:
\begin{itemize}
\item We first compute $\tilde{\xi}'=\sqrt{\xi_1\xi_2}$, choosing among
  the two roots the one which lies closer to $\xi_1$ and $\xi_2$.
\item We then compute the product of the $\frac{3V}{2}$ such defined
  $\tilde{\xi}'$ inside the unit circle and determine its phase
  $e^{i\delta}$.
\item we finally multiply each of the $\tilde{\xi}'$, both inside and outside
  the unit circle, by the same phase $e^{-i\frac{2\delta}{3V}}$, i.e.,
  \begin{equation}
    \label{eq:pairdeev}
   \tilde{\xi}=e^{-i\frac{2\delta}{3V}}\tilde{\xi}'=
   e^{-i\frac{2\delta}{3V}}\sqrt{\xi_1\xi_2}
   \,. 
  \end{equation}
\end{itemize}
The choice of the geometric mean automatically leads to satisfying
properties 1 and 2 of our list. At this stage property 3 will in
general not be satisfied, but it will after our global phase
correction. The phase of the averaged doublet is actually not
constrained by properties 1 and 2 if we choose it in the same way for
$(\xi_1,\xi_2)$ and its symmetric partner $(1/\xi_1^*,1/\xi_2^*)$. For
sufficiently large volumes the phase correction is tiny, 
and it also disappears in the continuum limit when the doublets
become exactly degenerate. 

Having obtained the averaged eigenvalues
$\{\tilde{\xi}_i\}_{i=1,\ldots,3V}$, we define the rooted
  determinant as 
\begin{equation}
  \label{eq:root_det}
    \sqrt{\det(P-e^{\hat{\mu}})}_{\rm P} \equiv 
\prod_{n=1}^{3V} (\tilde{\xi}_n - e^{\hat{\mu}}) 
\,,
\end{equation}
where ``P'' stands for ``paired''. Equation ~\eqref{eq:mupartfunc} for the
rooted determinant becomes
\begin{equation}
  \label{eq:mupartfuncroot}
  \begin{aligned}
  \frac{Z(\hat{\mu})}{Z(0)} &=  
   e^{-\frac{3 V}{2} \hat{\mu} } \left\langle  \prod_{i=1}^{3 V}\frac{
  \tilde{\xi}_i - e^{\hat{\mu}}}{ \tilde{\xi}_i - 1 }\right\rangle_0 \\
&= e^{-\frac{3 V}{2}\hat{\mu} }\,\left\langle {\cal
    P}_{\tilde{\xi}}(e^{\hat{\mu}})\right\rangle_0 
 =e^{-\frac{3 V}{2}\hat{\mu} }\,{\cal P}(e^{\hat{\mu}})\,,
  \end{aligned}
\end{equation}
with ${\cal P}$ a polynomial in the fugacity. We can now solve for the roots of this
polynomial to determine the Lee-Yang zeros of the partition function.

Finally, we note that both the usual definition and our new definition
of the rooted determinant are nonanalytic in the gauge fields at
finite fixed $\mu$. The source of the nonanalyticity is however slightly
different. 
The eigenvalues of the reduced matrix are of course analytic
functions of the link variables. When rooting eigenvalue by eigenvalue
as in Eq.~\eqref{eq:oldrooting}, 
the nonanalyticity comes from the eigenvalues of the reduced
matrix crossing the branch cut of the square root on the negative real
axis. In the case of geometric matching, nonanalyticity comes from
the minimization procedure, with eigenvalues sometimes changing pairs
as the gauge fields vary. A further source of nonanalyticity is the
small phase correction to ensure property 3.
The crucial difference is that in the case of geometric matching we
can be sure that after integration over the gauge fields the 
resulting partition function -- being a polynomial -- is analytic in
$\mu$ on the whole complex plane. This very nice feature of our
approach is a consequence of the simple fact that the sum of
polynomials is again a polynomial. With our approach, one can therefore
take the continuum limit of the radius of convergence directly, without
encountering analyticity issues, which was one of our goals stated in the
Introduction.
 
Finally, we note that the method can be straightforwardly generalized to
the $N_f=1$ case, simply by performing the geometric matching of pairs twice
or by defining an objective function which searches for quartets directly instead of pairs.
Close to the continuum limit, this is expected to lead to a correct identification of the
taste quartets.
\section{Numerical results}
\label{sec:numres}

We have performed numerical simulations at $\mu=0$ using a tree-level
Symanzik improved gauge action for the gauge fields and $2+1$ flavors
of rooted 2-stout improved staggered fermions~\cite{Aoki:2006br} at
physical quark masses. 
We used lattices of temporal size $N_t=4$ and
spatial size $N_s=\{6,8,10,12\}$ at $\beta=\{3.32,3.33,3.34,3.35\}$,
corresponding to temperatures below and up to $T_c$. 
A chemical potential is then introduced only for the light quarks,
with $\mu_u=\mu_d=\mu_B/3$, $\mu_s=0$.
For each simulation point we
gather on the order of 20000 configurations, separated by 10 HMC trajectories
each. For each gauge
configuration we performed a full diagonalization of the reduced
matrix $P$, using the publicly available MAGMA linear algebra library
for GPUs~\cite{MR1484478}. This is the most computationally intensive step of the
analysis, as it roughly scales with $V^3=N_s^9$. We then followed with the geometric matching 
of the eigenvalues to compute the partition function at finite $\mu$ via 
the reweighting formula Eq.~\eqref{eq:mupartfuncroot}. 

\begin{figure}[t]
  \centering
    \rotatebox{-90}{\includegraphics[width=0.33\textwidth]{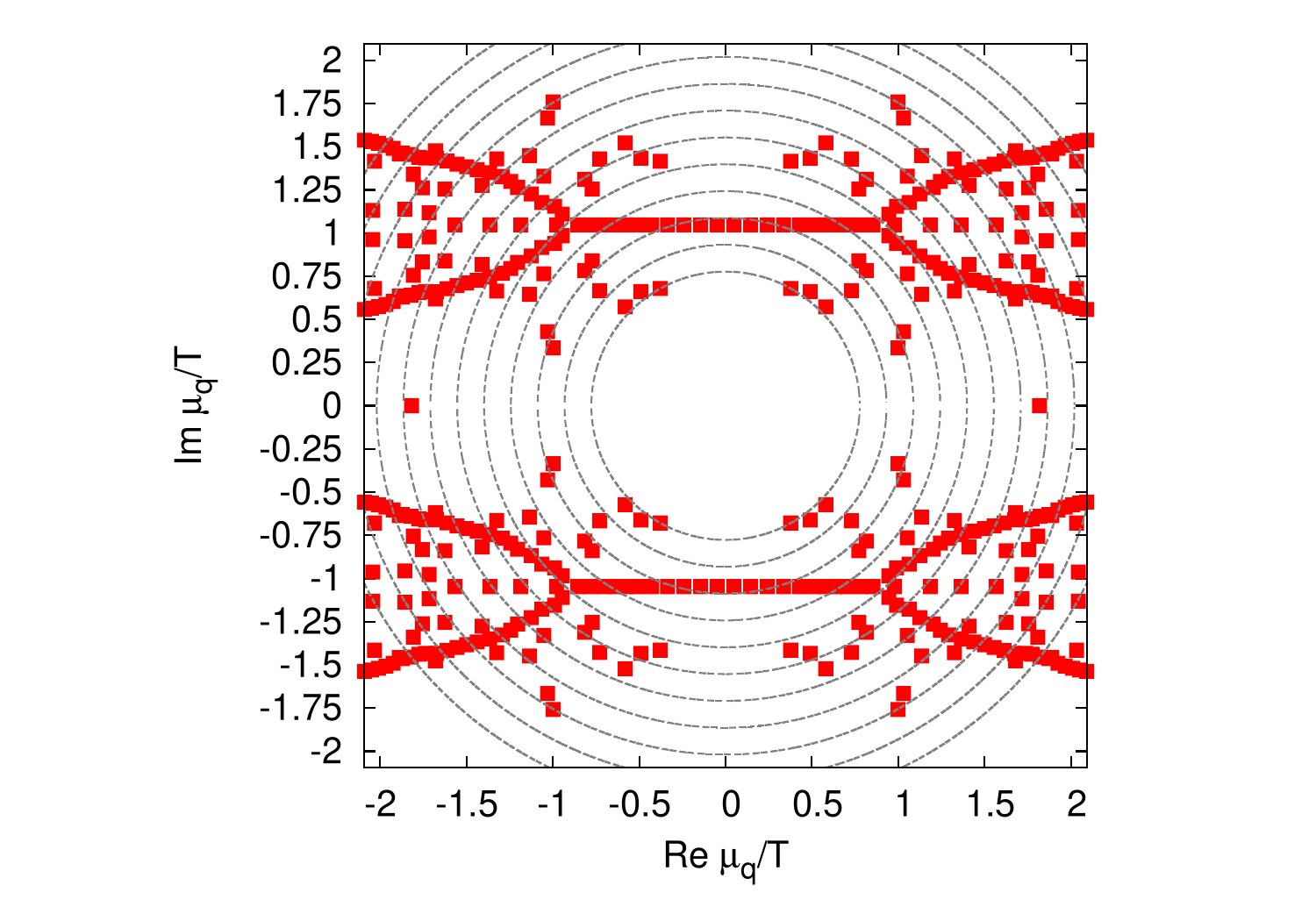}}
  \caption{Lee-Yang zeros in the complex $\mu/T$ plane. Here the lattice
    size is $12^3\times 4$ in lattice units and $\beta=3.35$. The 
    statistical errors are not shown.
    } 
  \label{fig:allLY}
\end{figure}

\begin{figure}[t]
  \centering
  \rotatebox{-90}{\includegraphics[width=0.33\textwidth]{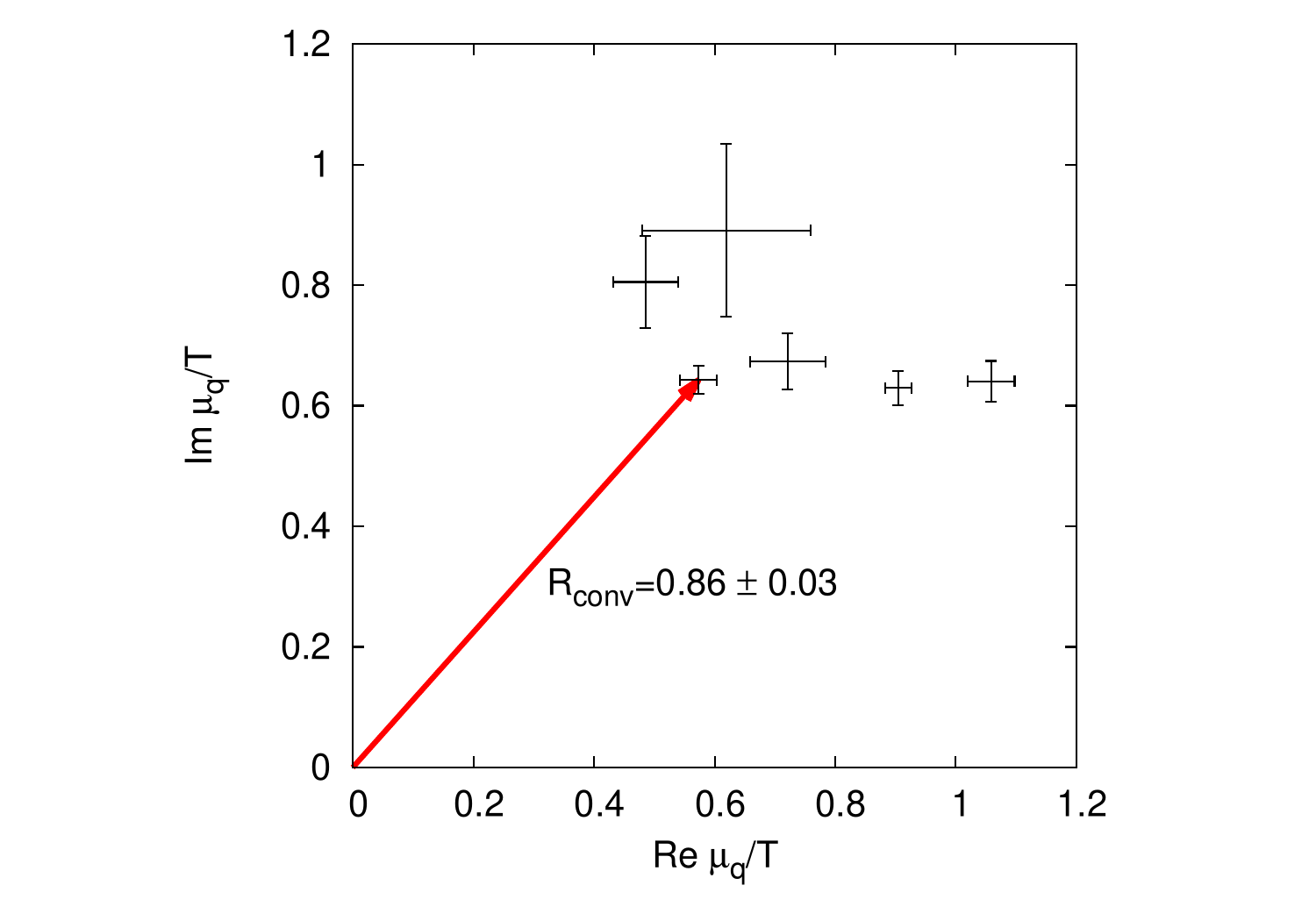}}     
  \caption{Radius of convergence on a $10^3\times 4$ lattice at $\beta=3.34$, together with the first few Lee-Yang zeros with error bars.}
  \label{fig:rcns10}
\end{figure}

\begin{figure}[t]
  \centering
  \rotatebox{-90}{\includegraphics[width=0.33\textwidth]{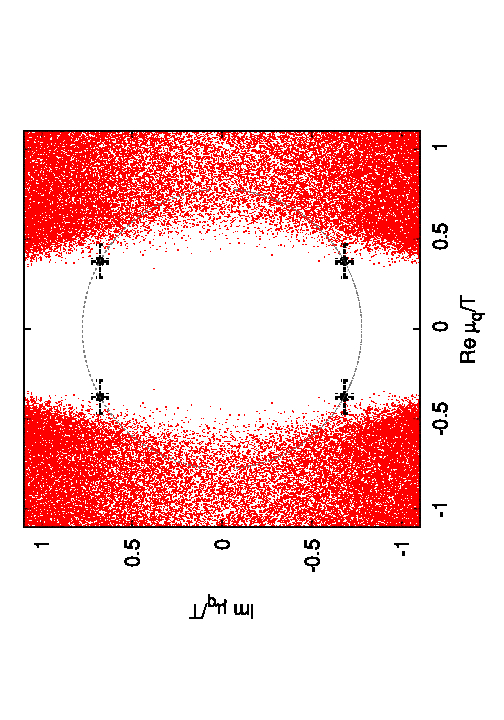}}    
    \caption{The logarithms of the eigenvalues of the reduced matrix $P$ for 3000 configurations superimposed on a single plot,
           as well as the leading Lee-Yang zeros determined with the new method 
					 for $\beta=3.35$ on a $12^3 \times 4$ lattice. The radius of convergence
           of the pressure defined with our method reaches inside the region inaccessible by the traditional definition.}
  \label{fig:LY0s_vs_spectral_gap}
\end{figure}

\begin{figure}[t]
  \centering
  \rotatebox{-90}{\includegraphics[width=0.33\textwidth]{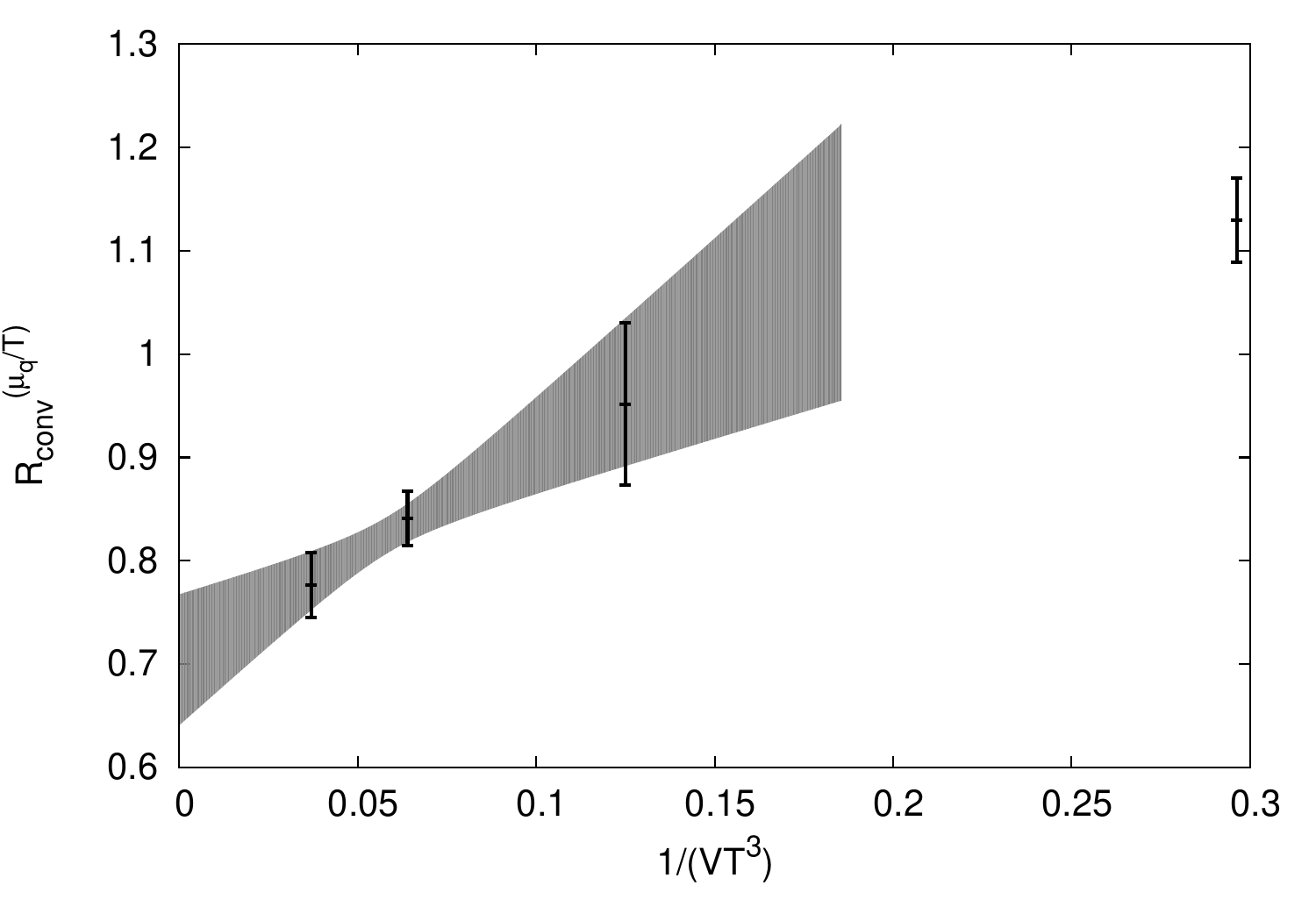}}    
  \caption{Radius of convergence against inverse volume at $\beta=3.35$. A linear fit in $1/V$ for the volumes $8^3, 10^3$ and $12^3$ is
    also shown.}
  \label{fig:rcnsV}
\end{figure}

\begin{figure}[t]
  \centering
  \rotatebox{-90}{\includegraphics[width=0.33\textwidth]{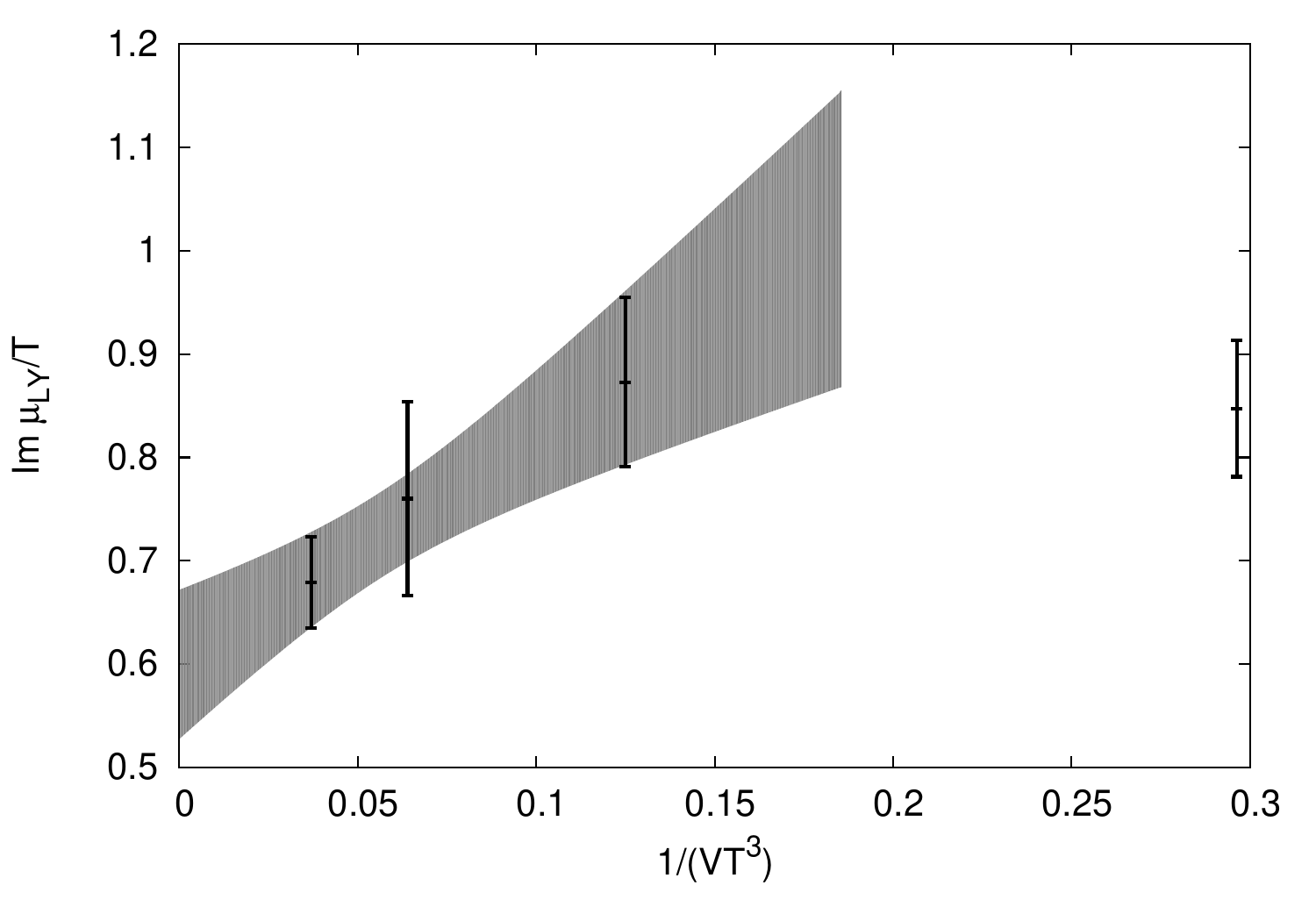}}    
  \caption{Imaginary part of the LY zero closest to the origin against
  inverse volume at $\beta=3.35$. A linear fit in $1/V$ for the
  volumes $8^3, 10^3$ and $12^3$ is 
    also shown.}
  \label{fig:immu}
\end{figure}

In order to determine $Z(\mu)/Z(0)$, for each configuration we
calculate the coefficients of the polynomial ${\cal
  P}_{\tilde{\xi}}(e^{\hat{\mu}})$ from the geometrically matched eigenvalues
$\tilde{\xi}_i$ using arbitrary precision arithmetic. 
Using Roberge-Weiss symmetry we can set to zero all
coefficients of the polynomial not of order $3n$. Using charge
conjugation symmetry instead we can set all coefficients to be real.
After averaging
the coefficients over gauge configurations, we determine all the roots
of ${\cal P}(e^{\hat{\mu}})$ using the Aberth method~\cite{Aberth},
which we also implemented in arbitrary precision. In
Fig.~\ref{fig:allLY} we show all the Lee-Yang zeros in the strip
${\rm Im}\,\mu\in [-\frac{2\pi}{3},\frac{2\pi}{3}]$ of the complex 
$\mu$ plane for a $12^3\times 4$ system at $\beta=3.35$. (Due to
rooting there are now half as many zeros in this strip than in the
unrooted case.) Error bars are not shown in Fig.~\ref{fig:allLY},
and while they are typically quite large, due to the sign problem,
the ones closest to $\mu=0$ zero have reasonably small errors. 
See Figs. \ref{fig:rcns10},\ref{fig:LY0s_vs_spectral_gap},\ref{fig:rcnsV} and \ref{fig:immu} (shown later).

\begin{figure}[t]
  \centering
  \rotatebox{-90}{\includegraphics[width=0.33\textwidth]{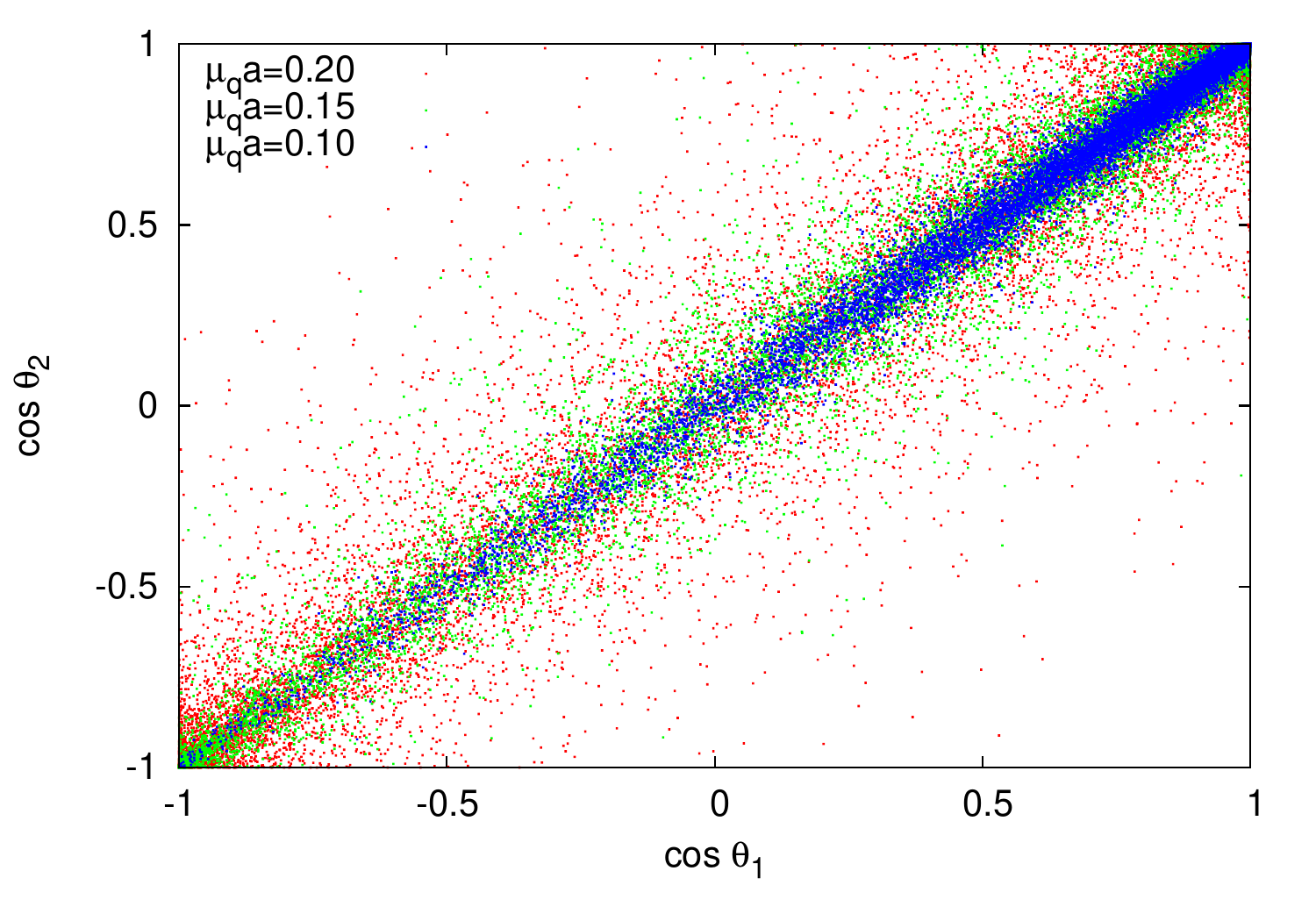}}
  \caption{Correlation of the phase of the determinant as obtained via
    geometric matching or by rooting separately the contribution of
    each eigenvalue, i.e. Eq.~\eqref{eq:oldrooting}, on a $12^3\times 4$ lattice at $\beta=3.34$. Here $\theta_1$ is the phase defined by the standard rooting procedure, while
    $\theta_2$ is the phase defined by our novel definition of the $N_f=2$ determinant.}
  \label{fig:cos}
\end{figure}

\begin{figure}[t]
  \centering
  \rotatebox{-90}{\includegraphics[width=0.33\textwidth]{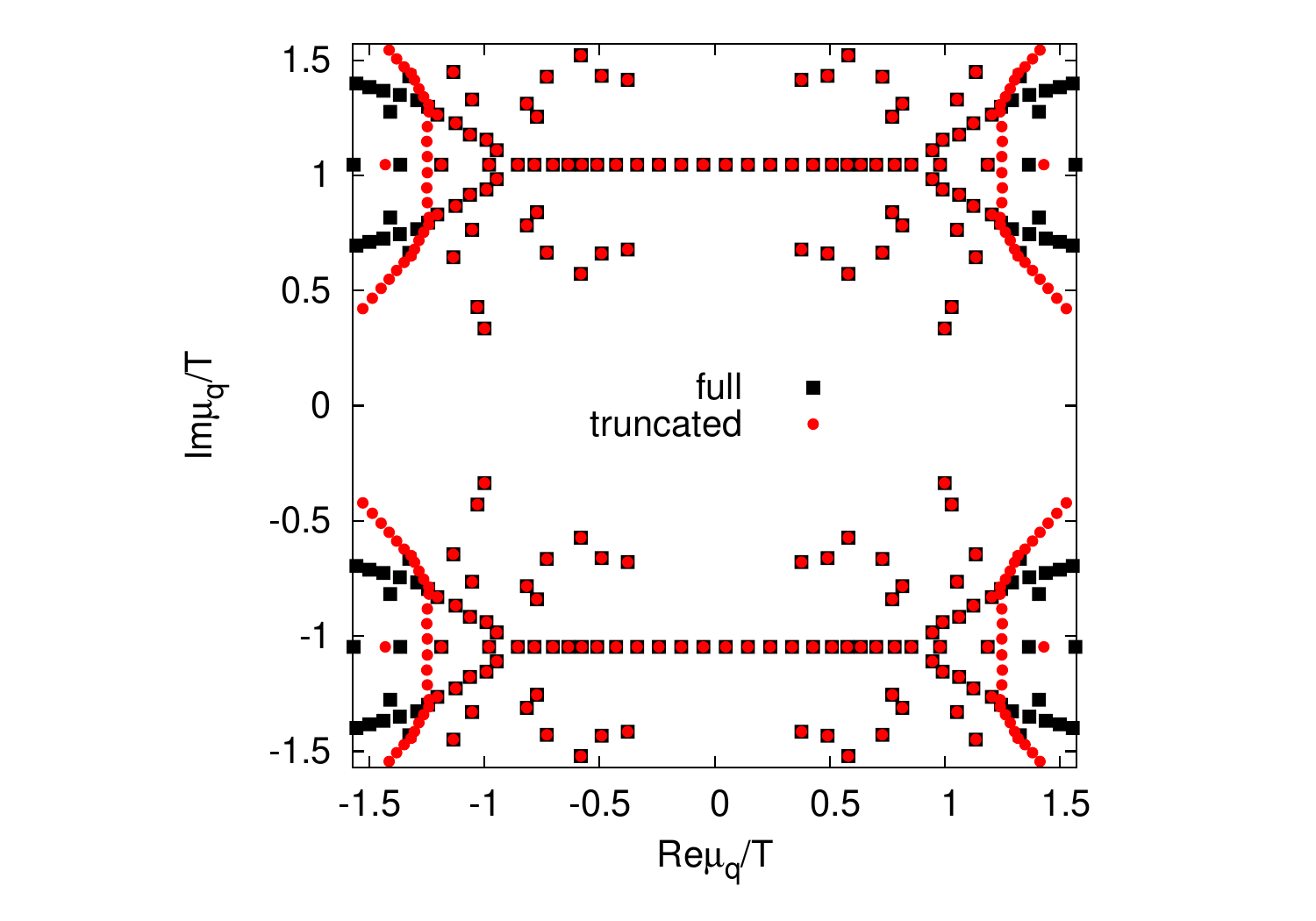}}
  \caption{Comparison between Lee-Yang zeros obtained with the full
    and the truncated polynomial, as described in the text. Here the
    lattice size is $12^3\times 4$ in lattice units and $\beta=3.35$. The 
    range of the ${\rm{Re}}\mu$ axis is chosen such that all roots of the truncated
    polynomial can be seen.
    }  
  \label{fig:trunc}
\end{figure}

As a first check of our method, we have studied the correlation
between the phase $\theta_1$ of the rooted determinant obtained via
the geometric matching procedure used in this paper and the phase
$\theta_2$ of the rooted determinant obtained by rooting each term of
the product separately, 
Eq.~\eqref{eq:oldrooting}.
The two methods are expected
to give the same continuum limit, but on a rather coarse lattice they
might be very different, leading to huge systematic uncertainties. 
In particular, a frequent relative 
change of sign would make the root of the fermionic
determinant a particularly ill-behaved quantity. In Fig.~\ref{fig:cos}
we plot the correlation between $\cos\theta_1$ and $\cos\theta_2$ for
three values of the chemical potential on a $12^3\times 4$ lattice at 
$\beta=3.34$. The two quantities are nicely positively correlated,
especially at small real $\mu$, indicating that the two methods will
give similar results on the real axis, even at a finite lattice spacing.

A possible systematic effect to take into account comes from the
relatively poor determination of the high-order coefficients of ${\cal
  P}$. While all the exact coefficients of the polynomial must be
positive, they can turn out to be negative on a finite sample due to
limited statistics and the sign problem. In such cases, while the
average is negative, the statistical error on the coefficient is above
$100\%$,  making them  
consistent with zero. We have then
checked the  
zeros in Fig.~\ref{fig:allLY} against those obtained truncating the Lee-Yang (LY) polynomial, removing
those terms for which we obtained a numerical estimate of the
coefficient that is compatible with zero. The comparison is shown in
Fig.~\ref{fig:trunc}: the physically relevant LY zeros near $\mu=0$
are unaffected by the truncation. This is true also for the zeros at
${\rm Im}\,\mu = \pm\frac{\pi}{3}$, corresponding to the thermal cut
of a fermion gas~\cite{Skokov:2010uc}. Notice that the unphysical zero at real $\mu$ visible in
Fig.~\ref{fig:allLY} is not in the ``safe'' region, and turns out to
be a numerical artifact. 

The LY zero closest to the origin determines the radius of convergence of
a Taylor expansion of $\log Z(\hat{\mu})$ around $\mu=0$. In
Fig.~\ref{fig:rcns10} we show the LY zeros closest to the origin,
including their error bars, on a $10^3\times 4$ lattice at
$\beta=3.34$, from which the radius of convergence is easily determined.

We also show how the Lee-Yang zero determined with our method compares to the 
spectral gap of the reduced matrix in Fig.~\ref{fig:LY0s_vs_spectral_gap}.
The radius of convergence of the pressure defined with our method reaches inside 
the region inaccessible by the traditional definition.

\begin{figure}[t]
  \centering
  \rotatebox{-90}{\includegraphics[width=0.33\textwidth]{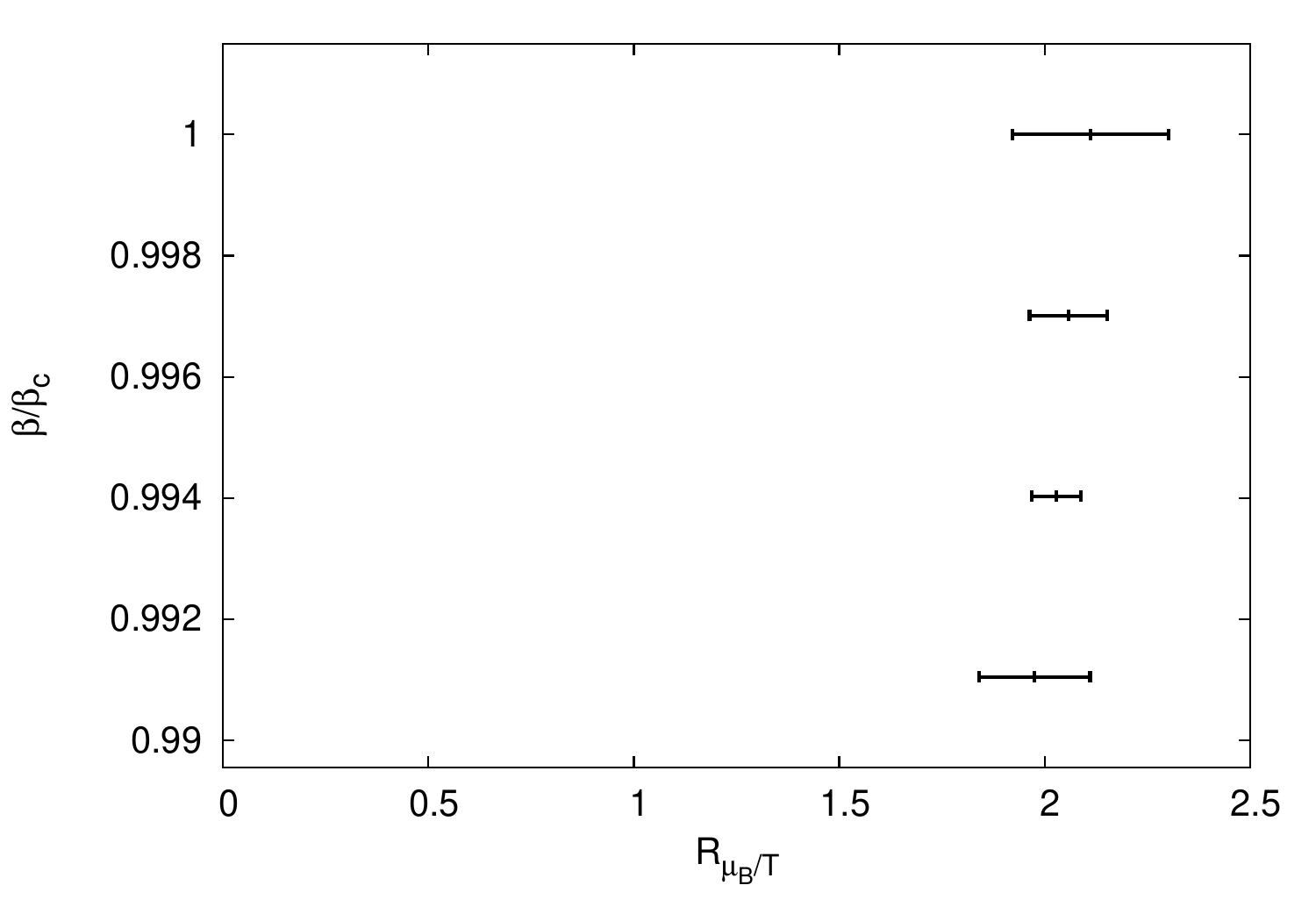}}     
    \caption{Radius of convergence at $V=\infty$ against temperature (bare gauge coupling), estimated by a linear fit in $1/V$ on the $8^3, 10^3$ and $12^3$ lattices.} 
  \label{fig:rcnsbeta}
\end{figure}

We have then repeated the procedure for all the available volumes.
The imaginary part of the leading Lee-Yang zero in general is expected to scale as
\begin{equation}
\label{eq:infvol}
    \left| {\rm Im} \mu \right| \sim a + \frac{b}{V^c} \rm{,}
\end{equation}
where for a first-order phase transition $a=0$ and $c=1$, while 
for a second-order phase transition $a=0$ and $c<1$ is given by the
critical exponents of the theory~\cite{Itzykson:1983gb, Deger:2019mgo}. In the absence of a phase transition $a>0$ and $c$ is in general not known.
Our data are not consistent with a first-order transition, neither it 
is consistent with a second-order transition in the 3D Ising or O(4) 
universality classes. At the moment our data is not precise enough to fit for all of $a,b$ and $c$. Empirically we find that the 
infinite volume extrapolations with $c=1$ lead to good fits. In this first study, we hence use a linear function 
in $1/V$ with the free parameters $a$ and $b$. We will also extrapolate the radius of convergence itself with the same ansatz.
 In Fig.~\ref{fig:rcnsV} we
illustrate this for the case $\beta=3.35$. 
The results for the extrapolated radius of convergence in the
thermodynamic limit for the various temperatures investigated in this
work are collected in Fig.~\ref{fig:rcnsbeta}.
The radius of convergence is $\mu_B/T \approx 2$ and almost constant in the range
of temperatures investigated in this paper. 
Note that this radius is larger than the radius of $\frac{3}{2} \frac{m_\pi}{T} \approx 1.4$ where reweighting from
the phase quenched theory is expected to break down due to the onset of pion condensation.

The existence of a genuine phase transition in the thermodynamic limit
is signaled by the vanishing of the imaginary part of the Lee-Yang
zero closest to the real axis. In Fig.~\ref{fig:immu} we show the
imaginary part of the LY zero closest to the origin as a function of
the volume at $\beta=3.35$. This extrapolates to a finite value,
indicating that the radius of convergence is not determined by a phase
transition. Also this value turns out to be almost
temperature independent in the range of temperatures considered here.

\section{Summary and outlook}
The radius of convergence of the Taylor expansion in $\mu_B$ is one of 
the most sought-after quantities in the finite-temperature QCD community~\cite{DElia:2016jqh,Vovchenko:2017gkg,Fodor:2019ogq,Giordano:2019slo,Savchuk:2019yxl,Mukherjee:2019eou}. 
The main reasons are that it gives a lower bound on the location of the 
critical end point, and also that it gives us insight in how far one can trust the 
equation of state calculated from a Taylor expansion around $\mu_B=0$.
Even if the lower bound on the location of the CEP happens to be not very stringent, this
validity region is still important, since viscous hydrodynamic simulations of heavy ion collisions
usually explore a very wide range of temperatures and baryochemical  potentials~\cite{Bass:2012gy}.

In this paper we have made two suggestions about how to obtain a reliable estimate
of this quantity in the framework of reweighting methods using the reduced-matrix 
approach~\cite{Hasenfratz:1991ax,Barbour:1997ej,Fodor:2001au,Fodor:2001pe,Fodor:2004nz,Csikor:2004ik}.  
The first one concerns the definition of the rooted staggered 
determinant itself at finite $\mu_B$. 
We have proposed an alternative definition that 
takes care of the analyticity issues of the partition function by making it --- 
up to an exponential factor
---  a polynomial in the fugacity.
This involves a procedure we dubbed geometric
matching, where to define the $N_f=2$ determinant at finite $\mu_B$ we judiciously group eigenvalues of the reduced matrix in pairs, and 
substitute each pair with the geometric mean of its members. This follows the spirit of the proposal of Ref.~\cite{Golterman:2006rw}.
While on the coarse $N_t=4$ lattices used in our study there are no clear taste quartets or pairs yet, in the
continuum our procedure is expected to give a reasonable definition of rooted determinant
on each gauge configuration. 

Our definition allows one to calculate the radius of convergence at any finite lattice spacing and
later take the continuum limit, instead of having to 
take the continuum limit of the Taylor coefficients first. We believe this is beneficial,
since the strong correlation between the statistical errors of the high-order Taylor coefficients present
on a single ensemble are what makes possible the determination of the
radius of convergence in the first place~\cite{Giordano:2019slo}, and taking the continuum limit of the
individual Taylor coefficients first can potentially wash out these strong correlations.

The second suggestion concerns the way the radius of convergence is calculated numerically:
we have demonstrated 
that a brute-force calculation of the Lee-Yang polynomial via reweighting, and a brute-force 
calculation of its roots is possible, at least for small lattices.
This provides a direct determination of the radius of convergence without 
relying on a finite-order Taylor expansion. This second suggestion could also be used with other fermion discretizations,
like Wilson fermions, where the rooting ambiguity discussed in this paper is not present.

Our numerical results on $N_t=4$ 2-stout improved staggered lattices suggest a radius of
convergence of $\mu_B/T \approx 2$ at and slightly below $T_c$, for  $\mu_s=0$.
While extending such a study to finer lattices is certainly a challenge, we believe it is a 
challenge worth pursuing, based on the conceptual advantage of having an actual Lee-Yang polynomial
at a finite lattice spacing, instead of a function that is nonanalytic in dense regions of the complex
chemical potential plane, as is the case with the standard definition of staggered rooting.

While the most important shortcoming of the present work is the use of a rather coarse lattice,
there are further improvements that could be made in the future. For one, for a precision determination of the
radius of convergence our assumption of $c=1$ in Eq.~\eqref{eq:infvol} should be relaxed. This requires more statistics on 
the currently used volumes and also data on larger volumes. Another possible improvement is the implementation of the strangeness
neutrality condition, so that the choice of our $\mu_S$ more closely resembles that of heavy ion collision experiments. This requires generalizing our procedure to identify taste quartets of eigenvalues. The most
straightforward way to achieve this is to simply repeat the pairing procedure twice. This would allow for
an arbitrary choice of $\mu_s$, in particular $\left\langle S \right\rangle=0$ could also be implemented.

Our discussion on the radius of convergence with the usual staggered determinant suggests 
that high-order cumulants of the baryon number are quite sensitive to taste symmetry 
breaking. It is therefore also an important question for the future to what extent the 
Taylor coefficients in $\mu_B$ obtained with our new definition of the $N_f=2$ determinant 
differ from those obtained
with the standard definition, and which of the two has a better continuum scaling. 
Once the issue of the continuum limit is under control, the finite-volume scaling of the Taylor coefficients in the continuum is also an important future question.

\section*{Acknowledgements}
We thank Sz.~Borsanyi, Z.~Fodor, and K.~K.~Sza\-bo for useful discussions and 
a careful reading of the manuscript.
This work was partially supported by the Hungarian National
Research, Development and Innovation Office - NKFIH grant KKP126769 
and by OTKA under the grant OTKA-K-113034. A.P. is supported by 
the J\'anos Bolyai Research Scholarship of the Hungarian Academy of 
Sciences and by the \'UNKP-19-4 New National Excellence Program of 
the Ministry of Innovation and Technology.

\appendix*

\section{Properties of the reduced matrix}
\label{sec:app}
The reduced matrix is given in temporal gauge by Eq.~\eqref{eq:stagred2}.
Since
$\det P_i=1$ and $\det L=1$, it readily follows that $\det P=1$. Since
$B_i^\dag=B_i$, it follows that $P_i^\dag=P_i$. Moreover
$P_i^{-1}=-\Sigma_2 P_i \Sigma_2$, with $\Sigma_2$ the block matrix
version of the Pauli matrix $\sigma_2$.
Since $\Sigma_2$ commutes with $L$, it follows that
$P^{\dag\,-1}=\Sigma_2 P \Sigma_2$, so that the spectrum $\{\xi_i\}$
of $P$ must be symmetric under $\xi_i \to 1/\xi_i^*$. 
Due to the
strict positivity of the determinant of the staggered operator at
$\mu=0$ for nonzero quark mass $m$, $\det(P-1)> 0$, one has
$|\xi_i|\neq 1$ and so eigenvalues come in pairs $(\xi_i,1/\xi_i^*)$ with
the same phase and inverse sizes. Denoting 
with $d=\prod_{i,|\xi_i|< 1} \xi_i=\sideset{}{'}\prod_i \xi_i$ the
product of eigenvalues inside the unit circle, one has $1= \det P =
d/d^* = e^{2i{\rm Arg\, d}}$, so $d$ must be real. From positivity of
the determinant of the staggered operator at $\mu=0$ it follows
$0< \sideset{}{'}\prod_i (\xi_i-1)(1/\xi_i^*-1)=
d (-1)^{3V} \sideset{}{'}\prod_i (1-1/\xi_i)(1-1/\xi_i^*) =
d\,  \sideset{}{'}\prod_i |1-1/\xi_i|^2 \,,
$
and so $d>0$.

The spectrum can be computed explicitly for configurations in which
both $U_j(t,\vec{x})$ and the untraced Polyakov loops
$W(\vec{x})=U_4(N_t-1,\vec{x})$ are uniform and all 
commute with each other. 
It is then possible to diagonalize them
simultaneously with a time-independent gauge transformation that
preserves the temporal gauge choice. Let $U_j={\rm diag}(\phi_a^{(j)})$,
$W={\rm diag}(\varphi_a)$, with
$\sum_a\phi_a^{(j)}=\sum_a\varphi_a=0$, and denote  $B_i=B$. The
eigenvectors $\psi$ of $P$ can be factorized into a color part $\chi$,
a space part $f$, and a two-dimensional part $v$ corresponding to the
block structure. One has 
\begin{equation}
  \label{eq:appredmat2}
  \begin{aligned}
P\psi^{s\,u}_{a\vec{p}}(\vec{x}) &=
\xi^{s\,u}_{a\vec{p}}\psi^{s\,u}_{a\vec{p}}(\vec{x})\,,\\ 
  \psi^{s\,u}_{a\vec{p}}(\vec{x}) &= \chi_a f^s_{a\vec{p}}(\vec{x})
  v^{s\,u}_{a\vec{p}} \,,\\ 
  B_a f^s_{a\vec{p}}(\vec{x}) &=s \lambda_{a\vec{p}}
  f^s_{a\vec{p}}(\vec{x}) \,,\\ 
  B_a &= 2 \eta_4\left(am+ \textstyle\frac{1}{2}\sum_{j=1}^3\eta_j (e^{i\phi^{(j)}_a} T_j
    -e^{-i\phi^{(j)}_a} T_j^\dag)\right)\\
      \lambda_{a\vec{p}}&= 2 \sqrt{(am)^2 + \textstyle\sum_{j=1}^3\sin^2(\phi^{(j)}_a +
    p_j)}\,,\\
  f^s_{a\vec{p}}(\vec{x}) &=
      \textstyle\frac{1}{\sqrt{2}}(1+s\frac{B_a}{\lambda_{a\vec{p}}}) 
  \textstyle\frac{1}{\sqrt{V}}e^{i\vec{p}\cdot\vec{x}}\,,\\
  U_j\chi_a &= e^{\phi^{(j)}_a}\chi_a\,, \quad W \chi_a=
  e^{\varphi_a}\chi_a\,,\quad   (\chi_a)_i = \delta_{ai}\,, 
  \end{aligned}
\end{equation}
with $a=1,2,3$, 
$(N_s/2\pi) p_j=0,1,\ldots N_s-1$,
and $v^{s\,u}_a$
satisfies the eigenvalue equation
\begin{equation}
  \label{eq:appredmat4}
  \begin{pmatrix}
    s\lambda_{a\vec{p}} & 1 \\ 1 & 0
  \end{pmatrix} v^{s\,u}_{a\vec{p}} = \zeta^{s\,u}_{a\vec{p}} v^{s\,u}_{a\vec{p}}\,.
\end{equation}
One finds
\begin{equation}
  \label{eq:appredmat5}
  \begin{aligned}
    \zeta^{s\,u}_{a\vec{p}} &= \textstyle \frac{s\lambda_{a\vec{p}}}{2} + u
      \sqrt{1+\left(\frac{\lambda_{a\vec{p}}}{2}\right)^2}\,,&&& u &= \pm 1 \\ 
v^{s\,u}_{a\vec{p}}&= N_{a\vec{p}}
\begin{pmatrix}
  u\sqrt{u\zeta^{s\,u}_{a\vec{p}}} \\ \frac{1}{\sqrt{u\zeta^{s\,u}_{a\vec{p}}}}
\end{pmatrix}\,, &&&
N_{a\vec{p}}^{-1} &=\textstyle
\sqrt{2}\left[1+\left(\frac{\lambda_{a\vec{p}}}{2}\right)^2\right]^{\frac{1}{4}}\,,
  \end{aligned}
\end{equation}
from which it follows that
\begin{equation}
  \label{eq:appredmat6}
  \begin{aligned}
  \xi^{s\,u}_{a\vec{p}} &=
  -e^{i\varphi_a}(\zeta^{s\,u}_{a\vec{p}})^{N_t} 
      = \textstyle  
-e^{i\varphi_a}\left[\frac{\lambda_{a\vec{p}}}{2} + su
    \sqrt{1+\left(\frac{\lambda_{a\vec{p}}}{2}\right)^2}\right]^{N_t}\,.
  \end{aligned}
\end{equation}
For generic
values of $\phi^{(j)}_a$, due to the invariance under $p_j\to p_j +
\pi$, each $\lambda_{a\vec{p}}^2$ is eightfold degenerate, and so
$\lambda_{a\vec{p}}$ is fourfold degenerate. Since
$\xi^{s\,u}_{a\vec{p}}$ depends only on the product $su$, there is a
further twofold degeneracy, leading to octets rather than quartets of
eigenvalues. This extra degeneracy factor of 2 is expected also in 
the continuum, corresponds to particle-antiparticle symmetry, 
and is also observed in the original staggered 
operator~\cite{Bruckmann:2008xr}.

%


\end{document}